\documentclass[prb,twocolumn,aps,showpacs,fixfloats]{revtex4}
\usepackage{graphicx}
\usepackage{bm}
\usepackage{amsmath,amssymb}
\usepackage{subfigure}
\usepackage{float}
\usepackage{latexsym}
\usepackage{color}
\usepackage{enumerate}
\usepackage{pdfpages}
\usepackage{tikz}
\usepackage{hyperref}

\begin{document}
\newcommand{\s}{\scriptscriptstyle}
\newcommand{\uu}{\uparrow \uparrow}
\newcommand{\ud}{\uparrow \downarrow}
\newcommand{\du}{\downarrow \uparrow}
\newcommand{\dd}{\downarrow \downarrow}
\newcommand{\ket}[1] { \left|{#1}\right> }
\newcommand{\bra}[1] { \left<{#1}\right| }
\newcommand{\bracket}[2] {\left< \left. {#1} \right| {#2} \right>}
\newcommand{\vc}[1] {\ensuremath {\bm {#1}}}
\newcommand{\tr}{\text{Tr}}
\newcommand{\Trans}{\ensuremath \Upsilon}
\newcommand{\Refl}{\ensuremath \mathcal{R}}

\title{Loss of adiabaticity with increasing tunneling gap in non-integrable multistate
Landau-Zener models}

\author{Rajesh K. Malla   and M. E. Raikh}

\affiliation{ Department of Physics and
Astronomy, University of Utah, Salt Lake City, UT 84112}

\begin{abstract}
We consider the simplest non-integrable  model of  multistate Landau-Zener transition.
In this model two pairs of levels in two tunnel coupled quantum dots are swept
passed each other by the gate voltage. Although this $2\times 2$ model
is  non-integrable, it can be solved analytically in the limit when the
inter-level energy distance is much smaller than their tunnel splitting.
The result is contrasted to the similar $2\times 1$ model,  in which one
of the dots contains only one level. The latter model does not allow
interference of the virtual transition amplitudes, and it is exactly solvable.
In $2\times 1$ model, the probability for a particle,
residing  at time
$t\rightarrow -\infty$ in one dot, to remain in the same dot at
$t\rightarrow  \infty$ falls off exponentially with tunnel coupling.
By contrast, in $2\times 2$ model, this probability {\em grows} exponentially
with tunnel coupling. The physical origin of this growth is the formation of the
tunneling-induced collective states in the system of two dots.
This can be viewed as manifestation
of the Dicke effect.
%
% which is the manifestation
%of the Dicke effect.
\end{abstract}

\maketitle

\section{Introduction}
\begin{figure}
\label{f1}
\includegraphics[scale=0.32]{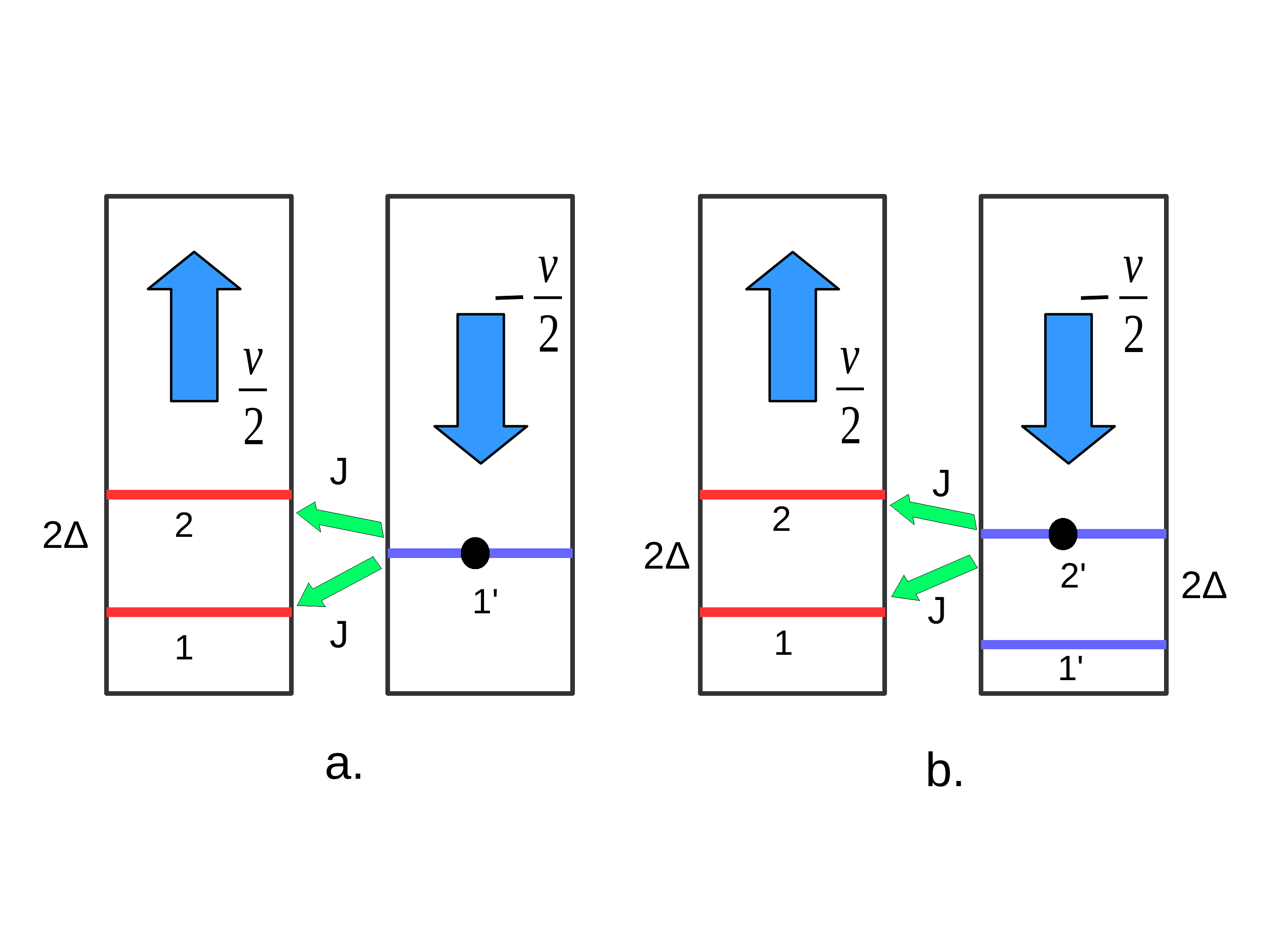}
\caption{(Color online)
Two elementary  multistate LZ models are illustrated;
(a) $2\times 1$ model of tunnel-coupled dots: a single level in the right dot is swept by two levels in the left dot with relative velocity, $v$; (b) $2\times 2$ model: the energy spacing, $2\Delta$, between the levels in both dots is the same. Each level in one dot is coupled with both levels in the other dot with coupling constant, $J$. }
\end{figure}
\begin{figure}
\label{f2}
\includegraphics[scale=0.12]{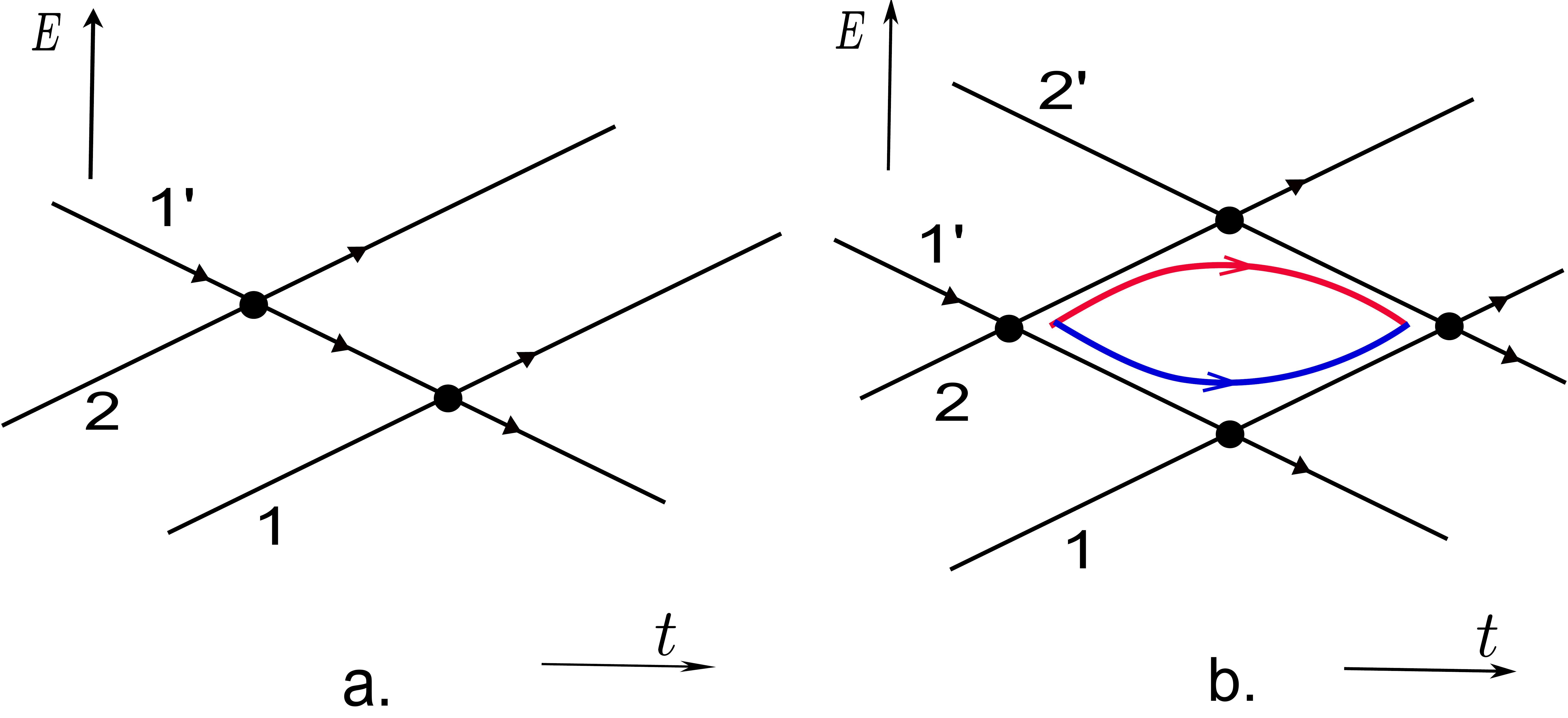}
\caption{(Color online) Different paths of the multilevel LZ transition are
illustrated for $2 \times 1$ (a) and $2\times 2$ (b) models. Red and blue
arrows illustrate the two interfering paths. }
\end{figure}
Motivations for the study of the transition probabilities
between multiple intersecting levels
(multistate Landau-Zener transitions)
were different over different periods of time.
Pioneering result for the full scattering matrix
was obtained  by Demkov and Osherov in Ref. \onlinecite{DemkovOsherov}
for a certain particular variant of level crossings.
The paper Ref.~\onlinecite{DemkovOsherov}
was motivated by the
research\cite{Demkov1964,Osherov1966,DemkovKomarov,Perel1971}
on inelastic atomic collisions.
Multilevel description of the electron transfer
in the course of the collision is required when
the crossing levels are dense, so that
the tunnel splitting exceeds the level spacing.
In this situation, the conventional Landau-Zener (LZ)
theory\cite{Landau1932,Zener1932,Majorana1932,Stueckelberg1932}
developed for a single crossing is inapplicable.

Later, the physics of multiple level crossings
had emerged in quantum optics\cite{Carroll},
in particular, in the problem of two optical transitions,
having a common level, in an atom driven by two laser beams.
Theoretical works of this
period\cite{bowtie1,BrundoblerImportant,DO1995,Usuki1997,Ostrovsky1997,DO2000,Ostrovsky2001,Sinitsyn2004,Shytov2004,Vitanov2005}
had broadened the class of exactly solvable models.
Also, for general mustistate models, the exact results
for certain elements of the scattering matrix
had been established.

Finally, the motivation for the very recent studies of the multilevel LZ
transitions\cite{Ashhab2014,Yuzbashyan2015,Ashhab2016,Sinitsyn2015,SinitsynJPhys2015,SinitsynLinChernyak2017,Sinitsyn2017}
was the ongoing experimental research on qubits manipulation
by time-dependent fields in relation to the information processing.
In these studies\cite{Ashhab2014,Yuzbashyan2015,Ashhab2016,Sinitsyn2015,SinitsynJPhys2015,SinitsynLinChernyak2017,Sinitsyn2017}
a number of new exactly solvable models were identified, although
the conclusion about their solvability was drawn on the basis of numerics.

The simplification, which allowed the authors of Ref. \onlinecite{DemkovOsherov}
to find the scattering matrix exactly, stemmed  from the assumption about
the time evolution of the energy levels. Namely,
it was assumed that $N-1$ out of $N$
levels evolved with the same velocity, and only one
level evolved with different velocity.
Thus, the number of crossings was $N-1$.
The behavior of the amplitudes to stay on
a given level at $t\rightarrow -\infty$,
i.e. far away from all crossings,   can be found semiclassically.
The contour integral method employed in Ref. \onlinecite{DemkovOsherov}
allows to establish the relations between
these amplitudes at $t\rightarrow -\infty$ and $t\rightarrow \infty$.
With $N-1$ crossings, these conditions are sufficient to fix
all $\frac{1}{2}\Big(N^2+5N-10\Big)$ nonzero
transition probabilities. The above approach has been employed
in all subsequent theoretical works with the exception of
Refs. \onlinecite{Usuki1997} and \onlinecite{Kayanuma1985}.
In these two  papers the transition probabilities were
derived upon summation of the perturbation expansion
in powers of the inter-level coupling strengths.

The fact that a given multistate LZ problem can be solved exactly
implies that the elements of scattering matrix can be constructed
from the partial LZ probabilities, $P_{\s LZ}$, for individual pairs
of intersecting levels. In other words, the time intervals  between
the successive intersections do not enter in the result even when these
intervals are much smaller than the characteristic time of LZ transition.
Yet another way to express this remarkable fact is that the independent crossing
approximation, valid for small tunneling gaps, remains applicable even when
the gaps are much bigger than the energy separation of the
neighboring crossing points.

Note that, for sufficiently slow drive velocities or for big enough LZ gaps,
when the individual $P_{\s LZ}$-values approach $1$, the ``survival"
probability for a particle to stay on the initial level
is exponentially small.
This immediately suggests that, for exactly solvable (integrable)
models, the survival probabilities fall off exponentially with increasing the gaps.
Then the question arises as to whether the above conclusion is valid
for non-integrable models.
This question is addressed in the present paper.  We focus on a simple example
of the electron transfer  between two multilevel quantum dots.
Our main finding is that the survival probability can, actually, {\em increase} with increasing
the tunneling gap. We relate this finding to the Dicke effect\cite{Dicke}.
The reason why the non-integrable model can be solved analytically
is that, for a very slow drive, the semiclassical approach for the time-dependent
amplitudes applies even in the vicinity of the LZ transition\cite{we1,we2}.

\section{The model}
\begin{figure}
\label{f3}
\includegraphics[scale=0.07]{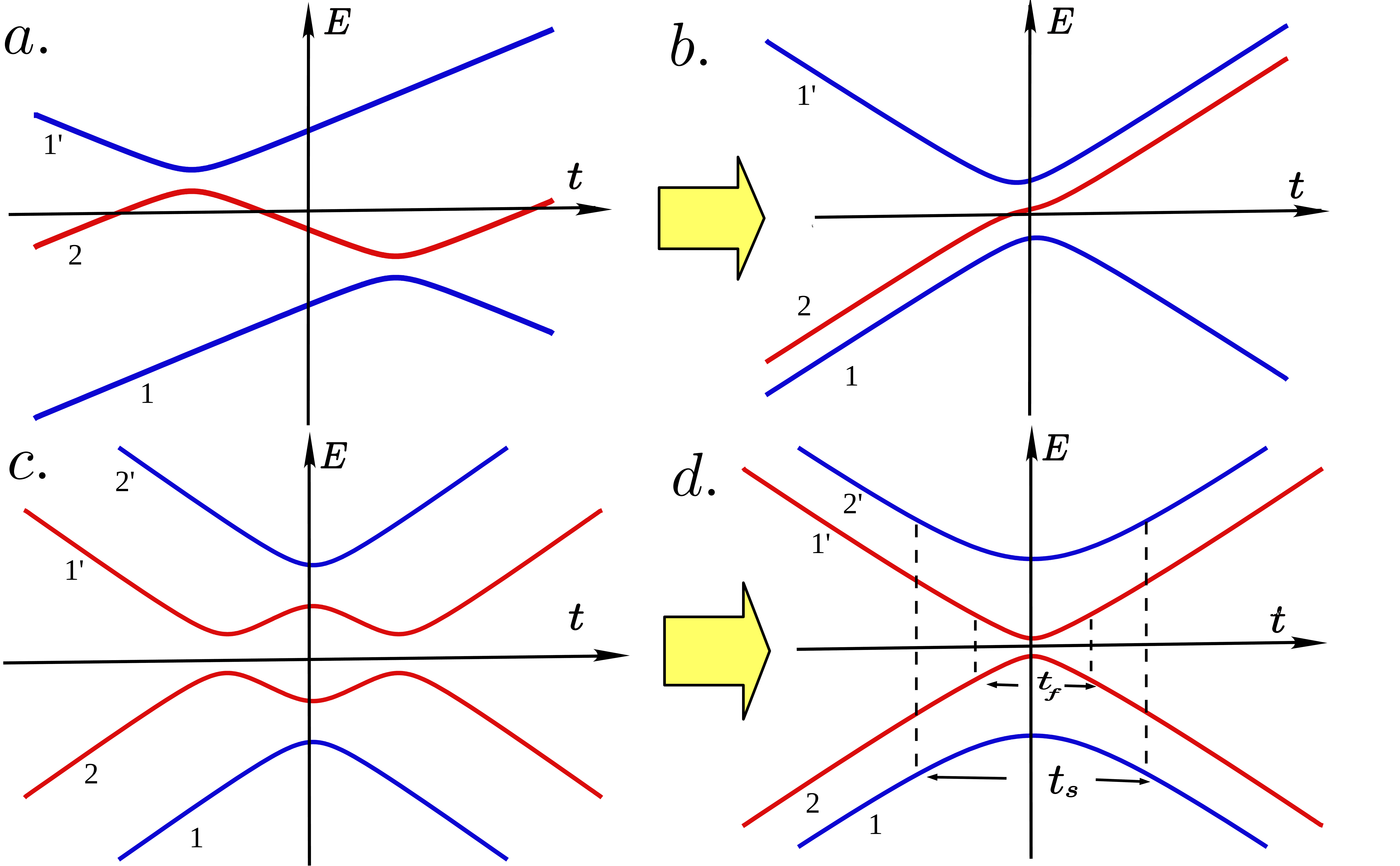}
\caption{(Color online) The evolution of the semiclassical levels in $2\times 1$ model (a), (b) and in $2\times 2$ model (c), (d) upon increasing the tunnel coupling. In $2\times 1$ model the two individual LZ transitions evolve into
a single transition at big $J$, while the middle branch (red) gets decoupled. In $2\times 2$ model the four individual LZ transitions evolve into the fast transition (red) and slow transition (blue). The levels are plotted in the units of $J$ versus the dimensionless  time, $vt/J$,
from the solutions of Eqs. (\ref{cubicequation})(a, b) and (\ref{Biquadratic4level})(c, d) for parameters $\Delta/J=10$ (a, c) and $\Delta/J=0.5$ (b, d). Vertical scale is set by the gap at $t=0$: $2\sqrt{2+\Delta^2/J^2}$ in (a, b), and $2(1+\sqrt{1+\Delta^2/J^2})$ in (c, d).}
\end{figure}

We start by illustrating  the difference between integrable and non-integrable models
using the simplest example of two quantum dots depicted in Fig. \ref{f1}. In Fig. \ref{f1}(a) there
 are two levels in the left dot separated by $2\Delta$ and one level in the right dot. The left-dot
 levels are driven, say, by the gate voltage, with velocity $v/2$, while the right-dot level is driven
 in the opposite direction with the same velocity. Both left-dot
 levels are coupled to the right-dot level by the same coupling constant, $J$.
 The matrix form of the Hamiltonian is the following:
\begin{equation}
\label{Matrixthreelevel}
\hat{H}_{2,1}=
\begin{pmatrix}
-\Delta - \frac{v t}{2} && 0 && J \\ \\ 0 && \Delta - \frac{v t}{2} && J \\ \\ J && J &&  \frac{v t}{2}
\end{pmatrix}.
\end{equation}
The evolution of the amplitudes, $a_1(t)$, $a_2(t)$, and $b_1(t)$, see Fig. \ref{f1}a, is
governed by the Schr{\"o}dinger equation
\begin{equation}
\label{3leveleigenequation}
i\begin{pmatrix}
\dot{a}_1 \\ \dot{a}_2 \\ \dot{b}_1
\end{pmatrix} = \hat{H}_{2,1} \begin{pmatrix}
a_1 \\ a_2 \\b_1
\end{pmatrix}.
\end{equation}
To find the semiclassical eigenvalues we substitute, $a_1(t), a_2(t), b_1(t) \propto \exp{\Big[i\int_{\s C}^t dt'\Lambda(t')\Big]}$, and arrive
to the following cubic equation for $\Lambda(t)$
\begin{align}
\label{cubicequation}
&\Lambda^3 +v t \Lambda^2  -\left(\Delta^2 +v^2 t^2 +2 J^2 \right)\Lambda  \nonumber \\
&= v t \left(-\Delta^2 +v^2 t^2 +2 J^2\right).
\end{align}
It is easy to see that the behavior of $\Lambda(t)$ (in the units of $J$) as a function of the dimensionless time, $vt/J$, is governed by a single dimensionless parameter $\Delta/J$. Upon changing this parameter, the semiclassical levels evolve
as shown in Fig. \ref{f3}. For small gap, $J \ll \Delta$, the levels exhibit two LZ transitions. At critical $\Delta =2^{1/2}J$ the slope of the middle level changes the sign. Finally, for large coupling, $J\gg \Delta$, the asymptotic solutions of Eq. (\ref{cubicequation}) are
\begin{equation}
\label{solutions}
\Lambda\approx vt,~~ \Lambda\approx \pm \left( v^2t^2+2J^2\right)^{1/2}.
\end{equation}
Eq. (\ref{solutions}) implies that in the limit, $\Delta \ll  J$, the middle semiclassical level
decouples from the upper and lower levels, which are given by the conventional LZ expressions with
$J$ replaced by $2^{1/2}J$.

The power of the integrability can be now illustrated as follows. Suppose that at $t=-\infty$ the
electron is in the right dot. For large $\Delta$, in order to remain in the right dot at $t\rightarrow \infty$,
it should survive two LZ transitions. Then the survival probability of each transition is given by
\begin{equation}
\label{standard}
Q_{LZ}\Big|_{\Delta \gg J}=\exp\left(-2\pi\frac{J^2}{v}\right).
\end{equation}
In the opposite limit of strong coupling the electron undergoes a single LZ transition. Integrability suggests
that the survival probability in this limit is given by the same formula as for the weak coupling, i.e. one should have
\begin{equation}
\label{relation}
Q_{LZ}\Big|_{\Delta \ll J}=\Big(Q_{LZ}\Big|_{\Delta \gg J}\Big)^2.
\end{equation}
Indeed, substituting $2^{1/2}J$, into Eq. (\ref{standard}), we realize that the relation Eq. (\ref{relation}) holds.

We now turn to non-integrable four-level model with the Hamiltonian

\begin{equation}
\label{Matrixfourlevel}
\hat{H}_{2,2}=
\begin{pmatrix}
-\Delta - \frac{v t}{2} & 0 & J & J \\ \\ 0 & \Delta - \frac{v t}{2} & J & J\\ \\ J & J & -\Delta + \frac{v t}{2}& 0\\ \\ J & J & 0 & \Delta + \frac{v t}{2}
\end{pmatrix}.
\end{equation}
In this model, there are two levels in the right dot, which are also split by $2\Delta$, see Fig. \ref{f1}b.
Instead of the amplitudes $a_1$, $a_2$, $b_1$, $b_2$, it is convenient to introduce the combinations
\begin{equation}
\label{A12}
A_1=a_1+a_2,~~~~~~A_2=a_1-a_2,
\end{equation}

\begin{equation}
\label{B12}
B_1=b_1+b_2,~~~~~~B_2=b_1-b_2,
\end{equation}
The time evolution of $A_1$, $A_2$, $B_1$, $B_2$ is governed by the system
\begin{eqnarray}
i{\dot A}_2-\frac{vt}{2}A_2-\Delta A_1=0,\label{S1}\\
i{\dot B}_2+\frac{vt}{2}B_2-\Delta B_1=0,\label{S2}\\
i{\dot A}_1-\frac{vt}{2}A_1-2JB_1=\Delta A_2,\label{S3}\\
i{\dot B}_1+\frac{vt}{2}B_1-2JA_1=\Delta B_2.\label{S4}
\end{eqnarray}
The equation for the semiclassical levels similar to Eq. (\ref{cubicequation}) takes the form
\begin{eqnarray}
\label{Biquadratic4level}
\Bigg[\left(\Lambda-\frac{vt}{2}\right)^2-\Delta^2\Bigg]
\Bigg[\left(\Lambda+\frac{vt}{2}\right)^2-\Delta^2\Bigg]\nonumber\\
=4J^2\Bigg[\Lambda^2-\left(\frac{vt}{2}\right)^2\Bigg].
\end{eqnarray}
The solutions of this equation are given by
\begin{equation}
\label{solutions4level}
\Lambda^2=2J^2+\left(\frac{vt}{2}\right)^2+\Delta^2 \pm 2\Bigg[J^4+J^2\Delta^2+\Delta^2\left(\frac{vt}{2}\right)^2\Bigg]^{1/2}.
\end{equation}
Our main point is that, in the limit of strong coupling, $J\gg \Delta$, the solutions Eq. (\ref{solutions4level}) can be
classified into ``slow" and ``fast", namely
\begin{eqnarray}
\Lambda_s \approx \pm \Bigg[4J^2+\left(\frac{vt}{2}\right)^2\Bigg]^{1/2},\label{slow4level} \\
\Lambda_f \approx \pm \Bigg[\frac{\Delta^4}{4J^2} +\left(\frac{vt}{2}\right)^2\Bigg]^{1/2}.\label{fast4level}
\end{eqnarray}
We see that, while the characteristic time for the slow solution is the conventional LZ time,
$t_s\sim J/v$, the characteristic
time for the fast solutions is $t_f\sim \Delta^2/Jv$, i.e. it is much shorter
(see also Fig. \ref{f3}d).
This is in  striking contrast with the integrable model.
Unlike integrable model, the splitting enters the result even if this splitting is very small.
Such a sensitivity to the times of the level crossings can be viewed as an indication that
it is interference of the scattering paths which makes the model non-integrable.
This interference is illustrated in Fig. \ref{f2}.

It is believed that in non-integrable models the two-level description is not applicable. In fact, Eq. (\ref{fast4level}) suggests that the scattering process decouples into two two-level LZ transitions with modified gaps. From Eqs. (\ref{slow4level}) and (\ref{fast4level}) we can readily
infer the survival probabilities of the slow and fast transitions:
\begin{eqnarray}
Q^{\s slow}_{LZ}&=\exp\Big[-2\pi\Big(\frac{4J^2}{v}\Big)\Big],\label{slowsurvival}\\
Q^{\s fast}_{LZ}&=\exp\Big[-2\pi\Big(\frac{\Delta^4}{4J^2v}\Big)\Big].\label{fastsurvival}
\end{eqnarray}
We see that, due to smallness of the  ``fast" gap, $Q^{\s fast}_{LZ}$ is much bigger than $Q^{\s slow}_{LZ}$, i.e. there is an anomalous survival of electron in a given dot. In other words, due to the interference, the adiabaticity of the transition between the two dots is lifted.

The above consideration was purely semiclassical. Thus, it applies when the probability
$Q^{\s fast}_{LZ}$ is small.
This requires that the splitting, $2\Delta$, while smaller than $J$,
exceeds $J\Big(v/J^2\Big)^{1/4}$, as follows from Eq. (\ref{fastsurvival}).
In the next section we go beyond the semiclassics and  demonstrate that
the condition of strong coupling, $J\gg \Delta$, is sufficient for Eq. (\ref{fastsurvival}) to apply.

\section{Anomalous survival probability}

It is seen from the system Eqs. (\ref{S1})-(\ref{S4}) that the amplitude $A_2$
and $B_2$, which are responsible for the fast LZ transition, are coupled indirectly
via $A_1$ and $B_1$. Respectively, the amplitudes  $A_1$ and $B_1$, responsible for the slow LZ transition, are coupled indirectly, via $A_2$ and $B_2$. The corresponding coupling constants are proportional to $\Delta$. On the other hand, the amplitudes $A_1$ and $B_1$ are coupled to each other directly with a coupling constant, $2J$, and this coupling is much stronger.
For this reason, we start with Eqs. (\ref{S3}), (\ref{S4})
and express $A_1$, $B_1$  via $A_2$ and $B_2$ in the following way
\begin{eqnarray}
\label{A1B1matrix}
\begin{pmatrix}
 A_1(t)\\ \\B_1(t)\end{pmatrix}
 =c_{s}^{+}(t) \begin{pmatrix} X_{s}^{+}(t)\\ \\Y_{s}^{+}(t)\end{pmatrix}
+ c_{s}^{-}(t) \begin{pmatrix} X_{s}^{-}(t)\\ \\Y_{s}^{-}(t)\end{pmatrix},
 \end{eqnarray}
where $X_{s}^{\pm}(t)$ and $Y_{s}^{\pm}(t)$ are the pairs  of the linear-independent solutions
of Eqs.   (\ref{S3}), (\ref{S4}) without the right-hand sides. In the presence of the right-hand side, in order to satisfy the system, the functions $c_{s}^{+}$, $c_{s}^{-}$ should
%satisfy
obey the following conditions
\begin{eqnarray}
&i\dot{c}_{s}^{+}X_{s}^{+}+i\dot{c}_{s}^{-}X_{s}^{-}=\Delta A_2, \label{condition1} \\
&i\dot{c}_{s}^{+}Y_{s}^{+}+i\dot{c}_{s}^{-}Y_{s}^{-}=\Delta B_2. \label{condition2}
\end{eqnarray}
Solving the system Eqs. (\ref{condition1}), (\ref{condition2}), we find
\begin{eqnarray}
&i\dot{c}_{s}^{+}=\frac{\Delta}{J W_s} \Big( A_2Y_{s}^{-} -B_2 X_{s}^{-}\Big) , \label{condition+} \\
&i\dot{c}_{s}^{-}= \frac{\Delta}{J W_s} \Big( B_2X_{s}^{+} -A_2 Y_{s}^{+}\Big), \label{condition-}
\end{eqnarray}
where we have introduced the notation
\begin{equation}
\label{Wronskian}
J W_s=X_{s}^{+}Y_{s}^{-}-Y_{s}^{+}X_{s}^{-},
\end{equation}
so that $W_s$ has a meaning of the Wronskian, which is time-independent.
Substituting Eqs. (\ref{condition+}, (\ref{condition-}) into Eq. (\ref {A1B1matrix}), and then
Eq. (\ref {A1B1matrix}) into Eqs. (\ref{S1}), (\ref{S2}), we arrive to the closed system
of integral-differential equations for $A_2(t)$, $B_2(t)$
\begin{widetext}
%\begin{eqnarray}
%i\dot{A}_2- \frac{vt}{2}A_2 +i \frac{\Delta^2}{J W_s}\int\limits_{-\infty}^{t}dt' A_2(t') \Big(
%X_{s}^{+}(t)Y_{s}^{-}(t')-X_{s}^{-}(t)Y_{s}^{+}(t') \Big)=i \frac{\Delta^2}{J W_s}\int\limits_{-\infty}^{t}dt' B_2(t') \Big(
%X_{s}^{+}(t)X_{s}^{-}(t')-X_{s}^{-}(t)X_{s}^{+}(t') \Big), \\
%i\dot{B}_2+ \frac{vt}{2}B_2 +i \frac{\Delta^2}{J W_s}\int\limits_{-\infty}^{t}dt' B_2(t') \Big(
%Y_{s}^{+}(t)X_{s}^{-}(t')-Y_{s}^{-}(t)X_{s}^{+}(t') \Big)=i \frac{\Delta^2}{J W_s}\int\limits_{-\infty}^{t}dt' A_2(t') \Big(
%Y_{s}^{-}(t)Y_{s}^{+}(t')-Y_{s}^{+}(t)Y_{s}^{-}(t') \Big),
%\end{eqnarray}
\begin{eqnarray}
&i\dot{A}_2- \frac{vt}{2}A_2 +i \frac{\Delta^2}{J W_s}\int\limits_{-\infty}^{t}dt' K_{xy}(t,t') A_2(t') =i \frac{\Delta^2}{J W_s}\int\limits_{-\infty}^{t}dt' K_{xx}(t,t') B_2(t'), \label{integraleq1}\\
&i\dot{B}_2+ \frac{vt}{2}B_2 -i \frac{\Delta^2}{J W_s}\int\limits_{-\infty}^{t}dt' K_{xy}(t',t)B_2(t') =i \frac{\Delta^2}{J W_s}\int\limits_{-\infty}^{t}dt' K_{yy}(t',t)A_2(t'),\label{integraleq2}
\end{eqnarray}
\end{widetext}
where the three kernels are defined as
\begin{eqnarray}
&K_{xx}(t,t')=X_{s}^{+}(t)X_{s}^{-}(t')-X_{s}^{-}(t)X_{s}^{+}(t'),\label{kxx}\\
&K_{yy}(t,t')=Y_{s}^{+}(t)Y_{s}^{-}(t')-Y_{s}^{-}(t)Y_{s}^{+}(t'),\label{kyy}\\
&K_{xy}(t,t')=X_{s}^{+}(t)Y_{s}^{-}(t')-X_{s}^{-}(t)Y_{s}^{+}(t').\label{kxy}
\end{eqnarray}
Up to now, we did not make use of the
smallness of $\Delta$. As we had found above, see Eq. (\ref{fast4level}), the characteristic
time of the fast LZ transition is $t_f\sim \Delta^2/Jv$, so that $vt_f \ll J$. This allows to
neglect the terms $\pm vt/2$ in the equations for $X_{s}$, $Y_{s}$, which, in turn, leads to
the following solutions
\begin{eqnarray}
X_{s}^{+}(t)&=\exp(2i J t),~~ Y_{s}^{+}(t)&=-\exp(2i J t), \label{x+y+} \\
X_{s}^{-}(t)&=\exp(-2i J t),~~ Y_{s}^{-}(t)&=\exp(-2i J t).\label{x-y-}
\end{eqnarray}
In fact, the true asymptotic behavior of the solutions Eqs. (\ref{x+y+}), (\ref{x-y-})
contains corrections originating from the $vt/2$ terms. For example, the asymptote for $X_s^{+}$
has the form
\begin{equation}
\label{trueasymptote}
X_{s}^{+}(t)=\exp(2i J t)+\exp\Big(-\pi \frac{4J^2}{v} \Big)\exp(-2i J t).
\end{equation}
The second term can be neglected due to the condition that the slow (not fast)
LZ transition is adiabatic.
Under this condition, the kernels also get greatly simplified, and acquire the form
\begin{eqnarray}
K_{xx}(t,t')&=2i\sin \Big[2J(t-t')\Big],\\
K_{yy}(t,t')&=-2i\sin \Big[2J(t-t')\Big],\\
K_{xy}(t,t')&=2\cos \Big[2J(t-t')\Big],
\end{eqnarray}
while the Wronskian assumes the value $JW_s=2$.
The above expressions for $X$ and $Y$ apply at short times $t \ll t_s\sim J/v$, i.e. at times shorter than the time of slow LZ transition.
Still, $t_s$ is much bigger than $t_f$, which allows to use the kernels
Eqs. (\ref{kxx})-(\ref{kxy}) in the system Eqs. (\ref{integraleq1}), (\ref{integraleq2}). The substitution yields
\begin{widetext}
\begin{eqnarray}
&i\dot{A}_2- \frac{vt}{2}A_2 +i \Delta^2\int\limits_{-\infty}^{t}dt' \cos\Big[2J(t-t')\Big] A_2(t') =\Delta^2\int\limits_{-\infty}^{t}dt' \sin \Big[2J(t-t')\Big] B_2(t'), \label{modifiedintegraleq1}\\
&i\dot{B}_2+ \frac{vt}{2}B_2 -i\Delta^2 \int\limits_{-\infty}^{t}dt' \cos\Big[2J(t-t')\Big]B_2(t') = \Delta^2\int\limits_{-\infty}^{t}dt' \sin \Big[2J(t-t')\Big]A_2(t').\label{modifiedintegraleq2}
\end{eqnarray}
\end{widetext}
As a next step, we argue that the kernels are rapidly oscillating functions, while $A_2(t')$ and $B_2(t')$ are slow functions of time. If we take them out of the integrals at $t'=t$, then the integral in the left-hand side will turn to zero, while the integral in the right-hand side will assume the value $1/2J$. As a result the system Eqs. (\ref{integraleq1}), (\ref{integraleq2}) will simplify to
\begin{eqnarray}
i{\dot A}_2-\frac{vt}{2}A_2=\frac{\Delta^2}{2J} B_2,\label{result1}\\
i{\dot B}_2+\frac{vt}{2}B_2=\frac{\Delta^2}{2J} A_2.\label{result2}
\end{eqnarray}
The above system describes the conventional LZ transition with coupling $\frac{\Delta^2}{2J}$, so that the corresponding survival probability will be given by Eq. (\ref{fastsurvival}).

In our derivation we did not assume that the fast LZ transition is adiabatic.
In fact, $Q_{LZ}^{fast}$ can be comparable to $1$. Certainly, the simplification of the integrals in Eqs. (\ref{modifiedintegraleq1}), (\ref{modifiedintegraleq2}) requires justification.
In the Appendix we consider this simplification in detail.

\section{Discussion}
In order to illuminate our main message, let us compare the theoretical
predictions for $2\times 2$ model in two limits: $\Delta \gg J$ and $\Delta \ll J$.
In the first limit the smallness of the LZ gap allows to obtain
the transition probabilities from simple reasoning.
Suppose that at $t=-\infty$ the electron is in the state $1$ in the left dot, see Fig. \ref{f2}.
In this situation, the survival implies that at $t\rightarrow \infty$ the electron remains in the state $1$, i.e. it survives two LZ transitions. The probability for this is $Q_{1\rightarrow 1}=Q_{LZ}^2$.
If at $t=-\infty$ the electron is in the state $2$, then the survival probability
is the sum of probabilities to remain either in the state $2$ or
in the state $1$. The first probability is  $Q_{2\rightarrow 2}=Q_{LZ}^2$. With regard to
the second probability, it should be taken into account that there are two
paths from $2$ to $1$, as illustrated in Fig. \ref{f2}. Corresponding
amplitudes interfere with each other. If the phase difference accumulated
during the time $2\Delta/v$ is random, one can add the corresponding probabilities,
so that $Q_{2\rightarrow 1}=2Q(1-Q)^2$. The average (with respect to the initial states)
survival probability reads
\begin{equation}
\label{probabilityinterference}
{\cal Q}_{L}
=\frac{Q_{1\rightarrow 1}+Q_{2\rightarrow 2}+Q_{2\rightarrow 1}}{2}=Q_{LZ}\left(Q_{LZ}^2-Q_{LZ}+1 \right).
\end{equation}
Consider now the limit $J\gg \Delta$. In $50$ percent of realizations the
electron at $t\rightarrow -\infty$ is in the state $a_1$ and in $50$ percent of realizations it is in the state $a_2$. The  slow and fast LZ transitions
take place within the states $A_1=\frac{1}{\sqrt{2}}(a_1+a_2)$ and  $A_2=\frac{1}{\sqrt{2}}(a_1-a_2)$, respectively. Averaging over the initial states suggests
that the $t\rightarrow -\infty$ probabilities to be in the states $A_1$ and $A_2$ are equal. This means that the average survival probability is  given by
\begin{equation}
\label{finalprobability}
{\cal Q}_{L}
=\frac{1}{2}\Bigg\{\exp\Bigg[-\frac{2\pi}{v}\frac{\Delta^4}{4J^2}\Bigg]
+\exp\Bigg[-2\pi\frac{4J^2}{v}\Bigg] \Bigg\}.
\end{equation}
We see that for, $J\gg \Delta$, the probability Eq. (\ref{finalprobability})
is much bigger than Eq. (\ref{probabilityinterference}), which seems counterintuitive. Moreover,  for $J\gg\Delta$, ${\cal Q}_{L}$ increases with
increasing the tunneling, i.e. the adiabaticity of the multilevel LZ transition gets suppressed.

%{\bf Most relevant is Fig. 9 from PHY\LambdaICAL REVIEW A
%95, 012140 (2017)}

In this paper we have focused on a simplest example of non-integrable model,
crossing of two pairs of levels in the left and in the right dots. It would
be certainly interesting to establish how general is our conclusion about the
anomalous survival of electron in a given dot. We can go one step further and
generalize the model to the case when two groups of  $N$ levels in the left and in the right dot
cross each other. Two assumptions\cite{DO1995}: (i) all $N^2$ couplings are the same, and  (ii) the levels are aligned at $t=0$, greatly simplify the analysis. Namely, instead of
Eq. (\ref{Biquadratic4level}) we get the following generalized equation
\begin{equation}
\label{generalequation}
\Bigg[\sum\limits_{k=1}^N \frac{1}{\Lambda+\varepsilon_k -\frac{vt}{2}}\Bigg]\Bigg[\sum\limits_{p=1}^N \frac{1}{\Lambda+\varepsilon_p +\frac{vt}{2}}\Bigg]=\frac{1}{J^2}.
\end{equation}
In the limit $J\gg \varepsilon_k$, which we assumed throughout the paper,
the structure of the solutions is the following. One solution describes the fast transition.
Neglecting $\varepsilon_k$ in the denominators, we find
\begin{equation}
\label{Snslow}
\Lambda_N^{slow}=\pm\Big[ \Big(\frac{vt}{2}\Big)^2 +N^2J^2\Big]^{1/2}.
\end{equation}
The fact that $\Lambda_N^{slow}$ is much bigger than $\varepsilon_k$ justifies neglecting $\varepsilon_k$ in the denominators. The corresponding survival probability is
\begin{equation}
\label{Nsurvivalslow}
Q_{LZ}^{slow}(N)= \exp\Bigg[-\frac{2\pi N^2J^2}{v}\Bigg]    =\left(Q_{LZ}^{slow}\right)^{N^2}.
\end{equation}
This result should be contrasted to
\begin{equation}
\label{Nsurvivalslow1}
 Q_{LZ}^{slow}(N)= \exp\Bigg[-\frac{2\pi NJ^2}{v}\Bigg]    =\left(Q_{LZ}^{slow}\right)^{N},
 \end{equation}
which emerges within the independent crossing approach and also applies to the integrable models. Indeed, to enforce integrability in a multilevel model, see e.g. Ref. \onlinecite{Sinitsyn2017} a portion of tunnel couplings should be set to be zero.

The other $N-1$ solutions of Eq. (\ref{generalequation}) describe the fast LZ transitions.
The values of $\Lambda$ for these solutions are close to the values, $\tilde{\Lambda}_N$, for
which the sum, $\sum\limits_{k=1}^N (\Lambda+\varepsilon_k)^{-1}$, passes through zero. This emphasizes the role of interference in the formation of the fast transitions. Indeed, the eigenvector, corresponding to a given
${\tilde \Lambda}_n$, is composed of many levels.
If all $\varepsilon_k$ reside in the interval $\Delta$, then the estimate for $\tilde{\Lambda}_n$ is also $\Delta$, which is much smaller than $J$. To find corresponding survival probabilities we expand Eq. (\ref{generalequation}) near $\tilde{\Lambda}_n$. The linear terms proportional to $vt/2$ get canceled out and we obtain
\begin{equation}
\label{finalequation}
\left(\Lambda- \tilde{\Lambda}_n \right)^2-\left(\frac{vt}{2} \right)^2=\frac{1}{J^2\Big[ \sum\limits_{k=1}^N \frac{1}{(\tilde{\Lambda}_n+\varepsilon_k)^2}\Big]^2}.
\end{equation}
From here we find that the survival probability corresponding to a given $\tilde{\Lambda}_n$
\begin{equation}
\label{Nsurviavalfast}
Q_{LZ}^{fast}(N)=\exp\Bigg\{-\frac{2\pi}{v}\frac{1}{J^2\Big[ \sum\limits_{k=1}^N \frac{1}{(\tilde{\Lambda}_n+\varepsilon_k)^2}\Big]^2}\Bigg\}.
\end{equation}
All the terms in the sum, $\sum\limits_k ({\tilde \Lambda}_n+\varepsilon_k)^{-2}$,
are positive, and the value of the sum is determined only by the levels $\varepsilon_k$
closest to $-\tilde{\Lambda}_n$. The distance between these levels is $\sim \left(\Delta/N \right)$.
Thus, the sum can be estimated as $\Big(\frac{N}{\Delta}\Big)^2$. Finally, within a numerical factor in the exponent, we have
\begin{equation}
\label{Nsurviavalfastfinal}
Q_{LZ}^{fast}(N)=\exp\Bigg\{-\frac{2\pi}{v}\frac{\Delta^4}{N^4J^2}\Bigg\}.
\end{equation}
We conclude that, for the fast transitions, the survival probability grows rapidly with $N$.

To explain qualitatively the loss of adiabaticity with increasing the tunneling gap we draw the analogy between this effect and the Dicke effect\cite{Dicke} well known in optics. If two emitters are separated by
the distance much smaller than the emitted wavelength, the radiation lifetime
of the pair increases drastically. This is because the two eigenmodes of the
oscillating emitters are the symmetric and antisymmetric combinations of the
individual oscillations. The antisymmetric mode weakly overlaps with the emission field. Hence the long lifetime. In the model we considered, due to
tunneling, the correct  eigenstates of, say, the left dot are also $A_1=\frac{1}{\sqrt{2}}(a_1+a_2)$ and  $A_2=\frac{1}{\sqrt{2}}(a_1-a_2)$. The
gap for $A_1$ is twice the gap in the individual LZ transition, while the gap
for $A_2$ is suppressed and decreases with $J$. This is the origin of the anomalous survival.
The bigger is the number of levels in each dot, the less strict is the requirement that all tunnel couplings are the same.\cite{Mesosopic}

\section{Concluding remarks}

%{\bf Multistate Landau-Zener model (MLZM) has provided
%unusually many nonperturbative exact results. Some of them
%correspond to interactions among only a few states [15–17],
%others describe truly many-body mesoscopic dynamics [18].
%In terms of complexity, DO and bowtie models [10,12] stay
%somewhere in-between. It has always been puzzling why
%all such seemingly different systems produce simple final
%results with many common properties.}

%{\bf In Phys. Rev. A 95, 012140 (2017) formula Eq. (43) detailed analysis
%of Demkov-Osherov model. Starting from Eq. (66) the analysis of integrable vs. non-integrable models.}
%
%\vspace{6mm}

%{\bf It has been noticed previously
% that, surprisingly, all
%known exactly solvable models of type with a finite number
%of interacting states have exact solutions for the scattering
%matrix that coincide with the prediction of the independent
%crossing approximation. Moreover, all such solvable models
%have two properties
%(i) the absence of the dynamic phase effect on transition
%probabilities in the semiclassical framework, and
%(ii) the exact crossing of adiabatic energies at intersections
%of diabatic states (i.e., diagonal elements of the Hamiltonian)
%without direct couplings.
%(2015, Sinitsyn)
%}
%
%
%
%
%\vspace{10mm}
%\vspace{7mm}

(i) It is common to judge on whether or not the system with many degrees of freedom is integrable basing on numerically generated level statistics in a limited spectral interval, see e.g.
Ref. \onlinecite{Statistics}. If the statistics is Poissonian, the system can be decoupled
into individual ``blocks" which do not interact with each other. This is an indication that the
system is integrable. If, alternatively, the level statistics is Wigner-Dyson, different energy
levels repel each other, suggesting that the corresponding eigenstates ``know"  about the entire
system. Such a system is non-integrable. With regard to multistate LZ models, similar approach
has been employed in Refs. \onlinecite{SinitsynLinChernyak2017,Sinitsyn2017}. If the time evolution of the semiclassical levels exhibited avoided crossings, the model was judged to be  non-integrable.
Certainly, it is the interference of many partial amplitudes that is responsible for the level repulsion in many-body systems. Similarly, in non-integrable multilevel LZ models the time evolution between two distant level crossings allows more than one path.

(ii) Although the integrable  models in Refs. \onlinecite{Sinitsyn2015,SinitsynJPhys2015}
contain interfering paths, the parameters of these models are fine-tuned in such a way that
interference drops out from the final results.

(iii) We would like to emphasize that there is no smooth transition between
the integrable model \ref{f1}(a) and non-integrable model \ref{f1}(b). Even if we introduce asymmetry between the two dots by assuming that the levels in the right dot are separated by $2\Delta_1 \ll 2\Delta$, we will not emulate the
$2\times 1$ situation  by taking the limit $\frac{\Delta_1}{\Delta}\rightarrow 0$. The formal reason for this is that the level degeneracy in the right dot
will be lifted due to coupling of the degenerate levels via the left dot.
The width of the gap corresponding to the fast LZ transition, with asymmetric
spacings, takes the value $\Delta\Delta_1/2J$.

%(iii) {\bf For Discussion: In this regard, the simplest toy model considered in the present paper
%demonstrates how the non-integrability affects the probability that the multilevel LZ transition
%takes place.
%While in integrable system, the stronger are the couplings, the higher is the adiabaticity, it
%is opposite in non-integrable system: growth of coupling strengths leads to the growth of the survival
%probability, and thus, the loss of adiabaticity.}

%{\bf The power of the exact result of Demkov and Osherov can be illustrated
%on the example of a three-state model. Two levels in the left well are
%split by $2\Delta$ and move with velocity $v/2$, while a {\em single}
%level in the right well move with velocity $-v/2$. The result of Demkov and Osherov
%suggests that, if at $-\infty$ a particle is in the {\em lower} level of the left well, then the probability to find it in the right well at $t\rightarrow \infty$ is
%$P_{\s LZ}=1-\exp{\left( -2\pi J^2/v  \right)}$. If the particle is initially in the
%{\em upper} level of the left well, then the transition probability is $P_{\s LZ}^2$. The splitting, $2\Delta$, does not enter into the scattering matrix. This is despite
%the middle diabatic level undergoes the sign-change of the slope at $t=0$ when the
%splitting gets smaller than $\Delta =2^{1/2}J$. It is equally counterintuitive that
%the scattering matrix does not depend on the relation between the LZ transition time
%$J/v$ and the time $\left(2\Delta\right)^{-1}$ of beating between the split levels.}
%

\vspace{3mm}

\centerline{\bf Acknowledgements}

\vspace{3mm}

Illuminating discussion with V. L. Pokrovsky and E. G. Mishchenko are gratefully acknowledged.
The work was supported by the Department of
Energy, Office of Basic Energy Sciences, Grant No. DE-FG02-06ER46313.

\vspace{3mm}

\section{Appendix}

\vspace{2mm}

In this Appendix we explore the assumptions leading from the system Eqs. (\ref{modifiedintegraleq1} ), (\ref{modifiedintegraleq2}) to the system
Eqs. (\ref{result1}), (\ref{result2}). We will first assume that the system Eqs. (\ref{result1}), (\ref{result2}) applies, and use it to trace the above
assumptions.

Consider the integral in the right-hand side in Eq. (\ref{modifiedintegraleq2}). Upon performing the integration by parts twice, it
can be cast in the form
\begin{multline}
\label{A1}
\int\limits_{-\infty}^{t} dt' \sin\Big[2J (t-t') \Big]A_2(t')\\ = -\frac{1}{2J}A_2(t)-\frac{1}{4J^2}\int\limits_{-\infty}^{t} dt' \sin\Big[2J (t-t') \Big]\frac{\partial^2A_2(t')}{\partial t'^2}.
\end{multline}
It is now convenient to combine the left-hand side with the term containing second derivative in the right-hand side
\begin{multline}
\label{A2}
\int\limits_{-\infty}^{t} dt' \sin\Big[2J (t-t') \Big]\Bigg[A_2(t')+\frac{1}{4J^2}\frac{\partial^2A_2(t')}{\partial t'^2}\Bigg]\\ = -\frac{1}{2J}A_2(t).
\end{multline}
If the system Eqs. (\ref{result1}), (\ref{result2}) applies, the ${\partial^2A_2}/{\partial t'^2}$ can be expressed through $A_2$.
Substituting this expression into Eq. (\ref{A2}), we get
\begin{multline}
\label{A3}
\int\limits_{-\infty}^{t} dt' \sin\Big[2J (t-t') \Big]\Bigg\{A_2(t')\Big[1- \frac{\Delta^4}{16J^4}+i\frac{v}{4J^2}-\frac{v^2t^2}{16J^2} \Big]\Bigg\}\\
=\frac{1}{2J}A_2(t).
\end{multline}
Now we see that taking $A_2(t)$ out of the integral amounts to keeping only the first term in the square brackets. Indeed, the second term is much smaller than $1$ by virtue of the condition $\Delta \ll J$. The third term is much smaller than $1$, since the slow transition is adiabatic.
With regard to the fourth term, the characteristic  $t'$ is of the order of time of the fast LZ transition. If the fast transition is adiabatic, then $t'$ is of the order of $\Delta^2/Jv$, so
that the fourth term is of order of the second term. If the fast transition is non-adiabatic, then
$t'$ is of the order of $v^{-1/2}$. In this limit the fourth term is of the order of the third term. In both cases the terms which we neglected are small.  Similar consideration justifies the simplification of the other integrals.

\end{document}